\documentclass[conference,10pt]{IEEEtran}
\IEEEoverridecommandlockouts
\usepackage{cite}
\usepackage{amsmath,amssymb,amsfonts}
\usepackage{algorithmic}
\usepackage{graphicx}
\usepackage{textcomp}
\usepackage{multirow}
\usepackage{xcolor}
\usepackage{url}
\def\BibTeX{{\rm B\kern-.05em{\sc i\kern-.025em b}\kern-.08em
    T\kern-.1667em\lower.7ex\hbox{E}\kern-.125emX}}

\usepackage{algorithm,algorithmic}

\usepackage{physics,bbding}
\usepackage{color,hyperref}
\hypersetup{colorlinks,
            breaklinks,
            linkcolor=black,
            urlcolor=black,
            anchorcolor=black,
            citecolor=black}

\begin{document}

\title{Quantum Computing Methods for Supply Chain Management\\
}

\author{\IEEEauthorblockN{Hansheng Jiang$^*$}
\IEEEauthorblockA{ \textit{$^*$Dept. of Industrial Engineering}\\
\textit{\qquad and Operations Research}\\
\textit{University of California, Berkeley}\\
Berkeley, CA 94706, USA \\
hansheng\_jiang@berkeley.edu}
\and
\IEEEauthorblockN{Zuo-Jun Max Shen$^*{}^\dagger$}
\IEEEauthorblockA{
\textit{$^*$Dept. of Industrial Engineering}\\
\textit{\qquad and Operations Research}\\
\textit{University of California, Berkeley}\\
Berkeley, CA 94706, USA \\
\textit{$^\dagger$Faculty of Engineering and}\\
\textit{Faculty of Business and Economics}\\
\textit{University of Hong Kong}\\
{Hong Kong, China}\\
maxshen@berkeley.edu}
\and 
\IEEEauthorblockN{Junyu Liu$^\ddagger{}^\mathsection$}
\IEEEauthorblockA{\textit{$^\ddagger$Pritzker School of Molecular Engineering}\\
\textit{Kadanoff Center for Theoretical Physics }\\
\textit{The University of Chicago} \\
Chicago, IL 60637, USA}
\IEEEauthorblockA{
\textit{$^\mathsection$qBraid Co.} \\
Chicago, IL 60615, USA\\
junyuliu@uchicago.edu}
}

\maketitle

\begin{abstract}
Quantum computing is expected to have transformative influences on many domains, but its practical deployments on industry problems are underexplored. We focus on applying quantum computing to operations management problems in industry, and in particular, supply chain management.  Many problems in supply chain management involve large state and action spaces and pose computational challenges on classic computers. We develop a quantized policy iteration algorithm to solve an inventory control problem and demonstrative its effectiveness. We also discuss in-depth the hardware requirements and potential challenges on implementing this quantum algorithm in the near term. Our simulations and experiments are powered by \texttt{IBM Qiskit} and the \texttt{qBraid} system. 
\end{abstract}


\section{Introduction}

Recent surveys indicate promising prospects of quantum computing in many fields, for example, in finance \cite{orus2019quantum,herman2022survey} and quantum chemistry simulation \cite{bauer2020quantum}. In this work, we explore what advantages quantum computing may provide for addressing important problems in the field of operations management. 

Supply chain management is a central problem in the field of operations management. Supply chain management is a discipline studying the flow of goods or services through all the different stages of the process. From the early days when inventory decisions were manually documented with pen and paper, supply chains have undergone a major transformation  to be more automated and efficient, thanks to the advances of new technologies such as digitalization, software development, and more computing power. Therefore, it is natural to wonder what the technology of quantum computing may offer to supply chain management.  In particular, it is well known that the large decision spaces of inventory control problems brought major computational challenges to supply chain management, and some recent studies have explored using deep learning methods for supply chain management \cite{gijsbrechts2021can}.

In this work, we conduct a modest inquiry of the prospect of quantum computing on supply chain management by focusing on a classic inventory control problem. We describe special features of this inventory control  problem that make quantum computing a suitable tool. We also discuss the practical limitations of certain quantum approaches in terms of hardware requirements and error correction. We run numerical experiments on IBM Qiskit \cite{qiskit} and the qBraid system \cite{qbraid} to establish the validity.

\section{Inventory Control Problem}
We consider a classic periodic-review inventory control problem in supply chain management \cite{snyderfundamentals}. At each time period, the manager of a warehouse needs to decide on the number of product units to order. We assume that there is no lead time and the ordering cost is proportional to the number of units ordered. At each time period, the demand is stochastic following some probability distribution known to the manager. When there is no sufficient inventory, excess demand is lost. The manager needs to make ordering decisions to balance the ordering cost and lost sales cost in the long run.  

\begin{figure}[htbp]
\centerline{\includegraphics[width =0.45 \textwidth]{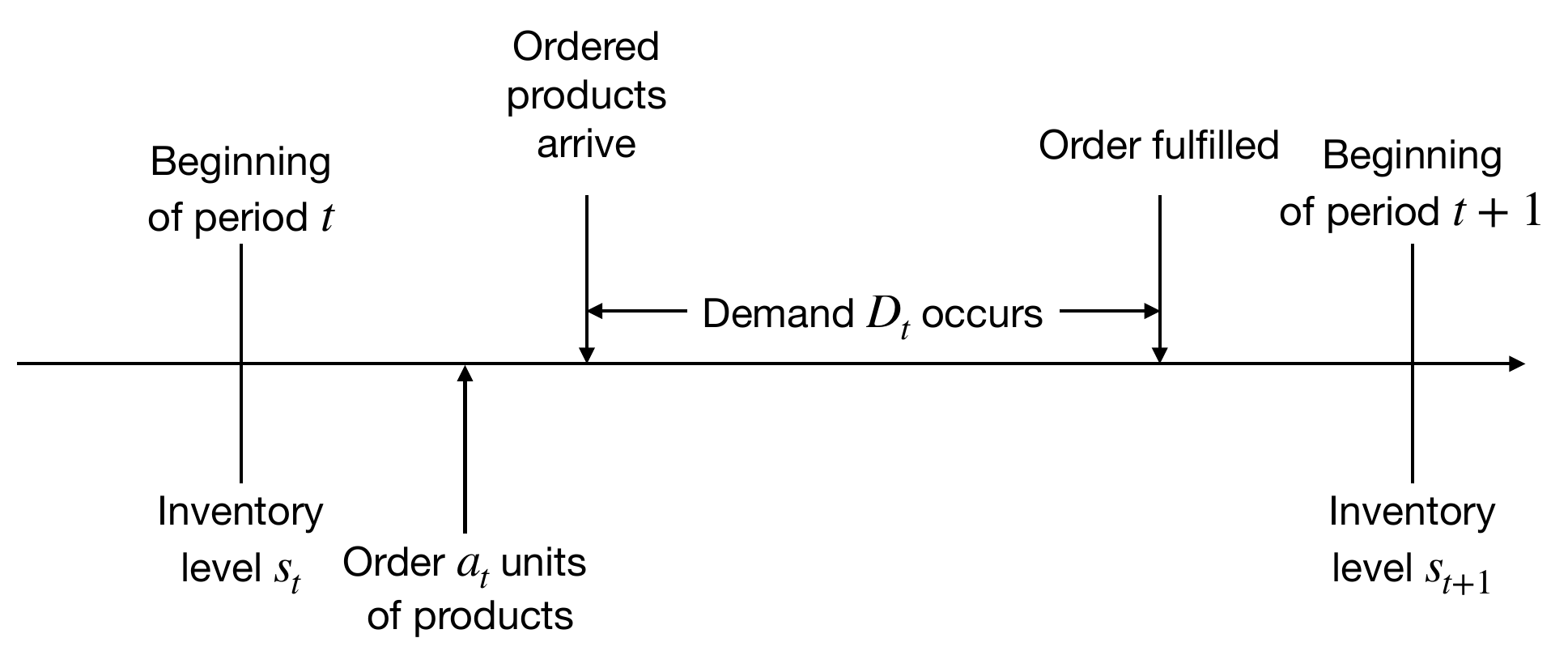}}
\caption{Sequence of events in the inventory control problem.}
\label{fig:event_illustration}
\end{figure}

This inventory control problem can be naturally modeled by the Markov Decision Process. Let $s_t$ represent the inventory level at the beginning of the $t$-th time period, and $a_t$ represent the number of units ordered by the inventory manager in the beginning of the period $t$. Correspondingly, the state space of $s_t$ and action space of $a_t$ are denoted by $\mathcal{S} = \{1,\dots, |\mathcal{S}|\}$ and $\mathcal{A} = \{1,\dots,|\mathcal{A}|\}$, respectively. Let $D_t$ represent the stochastic demand of this time period. We assume that the probability distribution of demand is stationary over time and can be specified by values $p_j = P(D_t = d)$ for $d = 1,2,\dots, D$ for some integer $D>0$. The inventory level in the beginning of $t+1$ , denoted by $s_{t+1}$, is captured by the following system transition function,
\begin{equation}
    s_{t+1} =  [s_t +a_t - D_t]^+,
\end{equation}
where $x^+:= \max\{x,0\}$. The sequence of events is illustrated in Fig.~\ref{fig:event_illustration}.

The reward of period $t$, denoted by $r(s_t,a_t)$, is given by
\begin{equation}
    r(s_t,a_t) = - h \cdot [s_t+a_t-D_t]^+- l \cdot [D_t-s_t-a_t]^+ ,
    \label{eq:reward_single}
\end{equation}
where $h$ is the unit holding cost, $l$ is the unit lost sales cost, and the negative signs are added because minimizing negative costs is equivalent to maximizing reward. Further, we use $r\in \mathbb{R}^{|\mathcal{S}||\mathcal{A}|}$ to denote the reward vector of a state-action pair, where $r(i,j)$ is given in \eqref{eq:reward_single}. 

We consider an infinite horizon discounted setting with some given discount factor $\gamma\in(0,1)$, and it is well-known that (see, e.g., \cite{puterman1994markov}) in this setting there exists a deterministic and stationary optimal policy $\pi:\mathcal{S} \rightarrow \mathcal{A}$ that maximizes the total discounted reward  over an infinite horizon. We follow the Q-function formulation \cite{agarwal2019reinforcement}, and the goal is to find policy $\pi$ that maximizes the Q-value function $Q^\pi(i,j)\in \mathbb{R}^{|\mathcal{S}||\mathcal{A}|}$ defined as 
\begin{equation}
Q^\pi(i,j) = \mathbb{E}\left[\sum_{t=0}^\infty \gamma^t r(s_t,a_t)|\pi,s_0 = i, a_0 = j  \right].
\end{equation}

We use $P^{\pi} \in \mathbb{R}^{|\mathcal{S}||\mathcal{A}| \times |\mathcal{S}||\mathcal{A}|}$ to denote the transition matrix on state-action pairs under  a stationary policy $\pi$, where the elements of $P^{\pi}$ are defined as 
\begin{equation}
P^{\pi}_{(i,j),(i',j')} = P(i'|i,j)\cdot \pi(j'|i'), \text{ for all } i,i' \in \mathcal{S}, j,j'\in\mathcal{A}.
\label{eq:p_pi_element}
\end{equation}
Under the notation of $Q^\pi$ and $P^\pi$, the Q-value function satisfies the following equation,
\begin{equation}
Q^\pi = r + \gamma P^{\pi} Q^\pi.
\label{eq:q_linear_system}
\end{equation}

\section{Quantized Policy Iteration Algorithm}

\subsection{Preliminaries on Policy Iteration}
A classic algorithm for solving the Markov Decision Process is the policy iteration  algorithm \cite{howard1960dynamic}, which generates a sequence of policies that gradually converge to the optimal solution over iterations. The policy iteration algorithm enjoys nice convergence properties and actually terminates within finite iterations. Specifically, it is proved in \cite{ye2011simplex} that the number of iterations of policy iteration can  be bounded in polynomial time $\tilde{\mathcal{O}}((|\mathcal{S}||\mathcal{A}|-|\mathcal{S}|)/(1-\gamma))$, where $|\mathcal{S}|$ and $|\mathcal{A}|$ stand for the size of state space and action space, respectively. However, each iteration of the policy iteration algorithm is time-consuming due to the sizes of the state space and the action space.  The per iteration complexity of the policy iteration algorithm has a time complexity of $|\mathcal{S}|^3+ |\mathcal{S}|^2|\mathcal{A}|$. It leaves space to leverage quantum computing to implement the policy iteration algorithm.

\begin{algorithm}[!ht]
 \caption{Policy Iteration Algorithm for Inventory Control}
 \begin{algorithmic}[1]
 \renewcommand{\algorithmicrequire}{\textbf{Input:}}
 \renewcommand{\algorithmicensure}{\textbf{Output:}}
 \REQUIRE demand distribution, cost parameters, $K$
 \ENSURE  optimal inventory reorder policy  \\
  \textit{Initialization} : initial policy $\pi_0$
  \FOR {$k = 0,1,\dots,K$}
  \STATE Calculate $Q^{\pi_k} \in \mathbb{R}^{|\mathcal{S}|}$ by a solving a linear system \begin{equation*}
Q^{\pi^k}  = (I - \gamma P^{\pi_k})^{-1} r
\end{equation*} \label{state:policy_evaluation}
  \STATE Obtain new policy $\pi_{k+1}$ by greedily solving the following maximization with respect to $v_{\pi_k}$: for all $i \in \mathcal{S}$ update 
\begin{equation*}
\pi_{k+1}(i) \in \arg\,\max_{j\in \mathcal{A}} Q^{\pi_k}(i,j)
\end{equation*}
\label{state:policy_improvement}
  \ENDFOR
 \RETURN $\pi_{K}$ 
\end{algorithmic} 
\label{alg:qpi}
\end{algorithm}

We have chosen the termination condition of Algorithm~\ref{alg:qpi} by setting the total number of iterations as some large enough number $K$ because it is known that policy iteration algorithm can converge in finite iterations \cite{ye2011simplex}. In Step~\ref{state:policy_evaluation} of Algorithm~\ref{alg:qpi},  we use quantum linear system solver to provide quantum states to approximate the Q function $Q^{\pi_k}$. Let $B_k$ denote $B_k:=I - \gamma P^{\pi_k}$, then the policy iteration step is finding the value of $B_k^{-1} r$, where the dimension of the matrix $B_k$ is $|\mathcal{S}||\mathcal{A}|$. In Step~\ref{state:policy_improvement}, the new policy $\pi_{k+1}(i)$  is obtained by finding the position of the largest element in  $Q^{\pi_k}(i,j)$ for the fixed $i$. 

A key observation is that the large matrices $B_k$ and $P^\pi_k$ have some sparsity properties for any deterministic policy $\pi$ under a very practical sparsity assumption on the inventory demand. Based on the definition in \eqref{eq:p_pi_element}, for every given $(i,j)$, $P^{\pi}_{(i,j),(i',j')}\neq 0$ only when $j' = \pi(i')$ and $P(i'|i,j)>0$. Assuming that the demand only takes $r\leq |\mathcal{S}|$ different values, then the number of non-zeros in $P^\pi_k$ is bounded by $r \times |\mathcal{S}||\mathcal{A}|$, and consequently the number of non-zeros in $B^\pi_k$ is bounded by $(r+1) \times |\mathcal{S}||\mathcal{A}|$, which is much smaller than the  total number of elements $|\mathcal{S}|^2|\mathcal{A}|^2$.

\subsection{HHL Algorithm for Policy Evaluation}

The Harrow-Hassidim-Lloyd (HHL) quantum algorithm \cite{harrow2009quantum} is a fundamental quantum computing algorithm for solving linear systems with exponential speed-up in theory.
For any matrix $B$,  HHL algorithm approximates the solution of $B^{-1} r$ in  $\mathcal{O}((\log(N) ) ^2)$ quantum steps where $N$ is the dimension of the matrix $B$ and $N = |\mathcal{S}||\mathcal{A}|$ in our case. Additionally, the time complexity also depends linearly on the sparsity $s$ defined as the number of non-zeros in $B$,  and $\kappa$ denoting the condition number of $B$. 

For notational simplicity, we consider the case of solving $Bq = r$, and the HHL algorithm works as follows. We first assume that $B$ is Hermitian, because otherwise we can consider the following equivalent formulation involved with a Hermitian matrix,
\[
\begin{pmatrix}
0 & B \\
B^\dagger & 0
\end{pmatrix}
\begin{pmatrix}
0\\
q
\end{pmatrix}
 = \begin{pmatrix}
 r\\
 0
 \end{pmatrix}.
\]
The vector $r$ is represented by a quantum state $\ket{r}$ with $\log_2N$ qubits, and the solution vector $q$ is also considered as a quantum state $\ket{q}$. Suppose $\ket{r} = \sum_l r_l \ket{E_l}$, where $\ket{E_l}$'s are the eigenvectors of $B$, and the corresponding eigenvalue of $\ket{E_l}$ is $\lambda_l$. Therefore, the equation $q = B^{-1}r$ can be encoded as
\begin{equation}
\ket{q} = \sum_l \lambda_l^{-1} r_l \ket{E_l}.
\label{eq:hhl_encode}
\end{equation}
Equation~\eqref{eq:hhl_encode} is achieved by first conducting quantum phase estimation to compute the eigenvalues $\lambda_l$, and then rotating an ancillary qubit of angle $\lambda_{\min}/\lambda_l$, where $\lambda_{\min}$ is the smallest eigenvalue, and lastly uncomputing the quantum phase estimation to obtain
\begin{equation}
    \sum_l r_l \ket{E_l} \left(\frac{\lambda_{\min}}{\lambda_l} \ket{1} + \sqrt{1 - \frac{\lambda_{\min}^2}{\lambda_l^2}} \ket{0} \right).
\end{equation}

While the $\mathcal{O}((\log N)^2)$ complexity is much faster than $\mathcal{O}(N\log N)$ of the classic algorithms and in particular it is an exponential speed-up, retrieving $x$ from $\ket{x}$ requires $\mathcal{O}(N)$ repetitions to get all $N$ components, and the cost of HHL in practice still remains prohibitive \cite{scherer2017concrete}.

\subsection{Variational Algorithms}
\label{subsec:variational}
The HHL algorithm is expected to be useful in the long term with fault-tolerant quantum computing technologies. However, in the near-term with the Near-term Intermediate Scale Quantum (NISQ) regime \cite{Preskill2018quantumcomputingin}, people are interested in finding variational analogs of HHL with similar capabilities in practice \cite{bravo2019variational,huang2021near}. Similar to the spirit of deep learning \cite{lecun2015deep}, those algorithms do not have a theoretically guaranteed computational advantage against their classic counterparts, but they might practically perform well in the near-term quantum devices \cite{cerezo2021variational}. 

More precisely, general variational quantum algorithms are specified by the following unitary operators,
\begin{align}
U = \prod\nolimits_{\ell  = 1}^L {{W_\ell }{U_\ell }}  = \prod\nolimits_{\ell  = 1}^L {{W_\ell }\exp (i{X_\ell }{\theta _\ell })}~,
\end{align}
where $W_\ell$'s are fixed unitary operators (constant gates in the quantum circuits), $X_\ell$'s are Hermitian operators (usually elements from the Pauli group), and $\theta_\ell$'s are variational classic parameters. The above unitary operations could be used to make loss functions based on the quantum measurements. Say that we initialize the quantum circuit by the state $\ket{\Psi_0}$, we could construct the following loss function,
\begin{align}
    \mathcal{L} = \bra{\Psi_0} U^\dagger  (\theta)O U \ket{\Psi_0}~.
\end{align}
The task now is to find the minimal eigenvalue of the Hermitian operator $O$, which might be obtained by traditional gradient descent algorithms,
\begin{align}
\theta_\ell (t+1)-\theta_\ell (t) = -\eta \frac{\partial \mathcal{L}} {\partial \theta_\ell}~,
\end{align}
with the learning rate $\eta$ and the number of iterations $t$. In supervised learning, we could encode supervised data in $\ket{\Psi_0}$. Thus, supervised learning could be performed with quantum machines similarly. 

One could give a variational version of the HHL algorithm by simply preparing variational states $\ket{x(\theta)} = U(\theta) \ket{\Psi_0}$ such that one could minimize the difference between  $A \ket{x(\theta)} $ and $b$, in order to solve the matrix inversion of $A$. We adapt the variational ansatz in \cite{bravo2019variational}, where people find a comparable performance against HHL with shorter circuit depth. We will primarily study the implementation of it in the near-term simulation and benchmarks.  

\section{QRAM Hardware Requirements}
When applying HHL-like algorithm, one of the primary challenges is to inform quantum computers the matrix elements of $A$ we want to invert. This problem could be solved, in principle, using so-called Quantum Random Access Memory \cite{giovannetti2008quantum} (QRAM). With the data size $N$, QRAM could use $\mathcal{O}(N)$ qubits, and $\mathcal{O}(\log N)$ time to implement the following unitary operator using a parallel way, 
\begin{align}
\sum_i \alpha_i \ket{i} \ket{0} \to \sum_i \alpha_i \ket{i} \ket{x_i} 
\end{align}
where $\alpha_i$'s are arbitrary coefficients and $x_i$'s are the data. In practice, large-scale, fault-tolerant QRAMs are challenging to build \cite{hann2021practicality,hann2021resilience,chen2021scalable}. However, it will be visionary to estimate how hard it is in practice to implement QRAM and HHL for given physical devices and algorithm requirements.

\begin{figure}[htbp]
\centerline{\includegraphics[width =0.45 \textwidth]{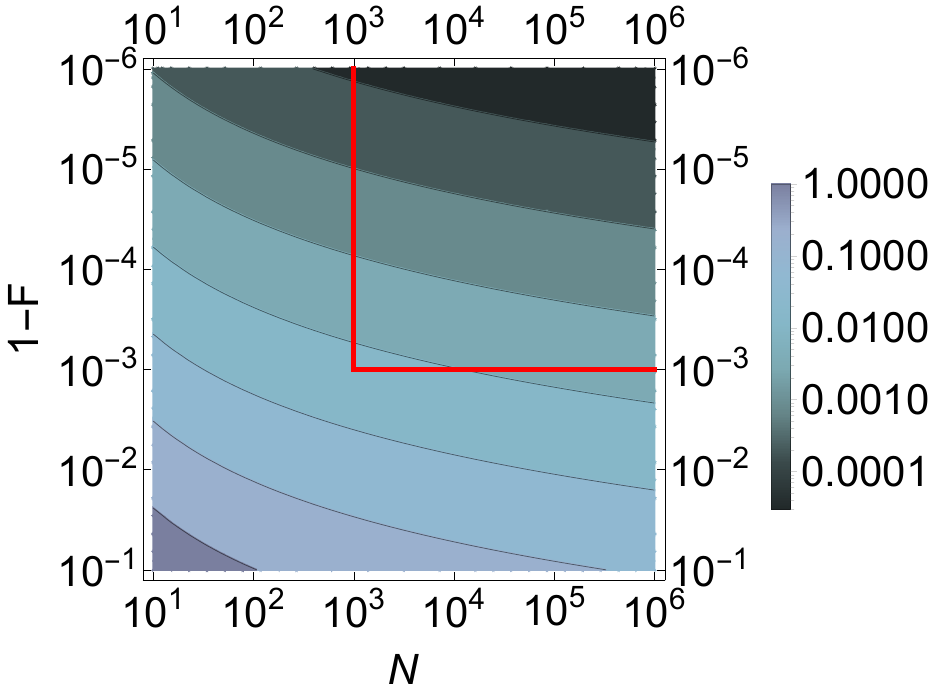}}
\caption{Constraints on the error rate $\varepsilon$ from the Markov Decision Process.}
\label{fig:qram1}
\end{figure}

\begin{figure}[htbp]
\centerline{\includegraphics[width =0.45 \textwidth]{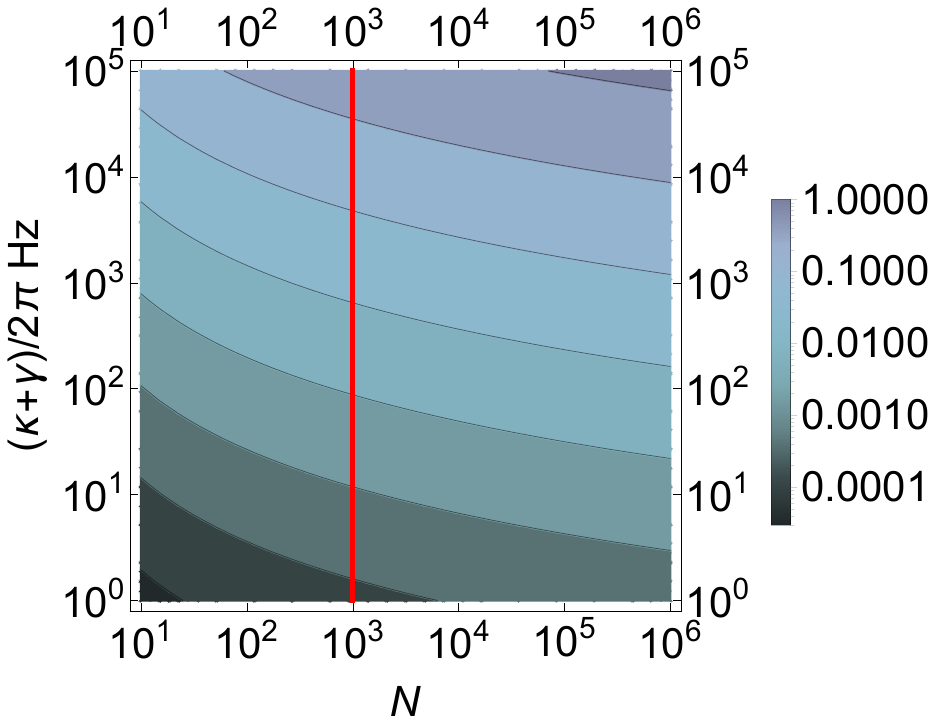}}
\caption{Constraints on the precision $1-F$  from the Markov Decision Process problem, with the decoherence rate $\kappa+\gamma$. }
\label{fig:qram2}
\end{figure}

In Figure \ref{fig:qram1} and \ref{fig:qram2}, we study how many physical parameters are needed for given requirements from current Supple Chain Management. Here, we make use of the hardware models of QRAM \cite{hann2021practicality,hann2021resilience}. Assuming that the size of the matrix is $10^3$, with the precision $10^{-3}$, which could typically happen in the current Supple Chain Management technology, we bound the required decoherence rate $\kappa+\gamma$ in the hybrid Circuit quantum electrodynamics system for realizing QRAM. A general conclusion is that it might be challenging to realize those requirements with the current physical devices, it will be helpful to study possibilities of realizing those experiments in the long term. 

Finally, we comment that the classic-quantum interfaces are essential for HHL-based algorithms. QRAM architectures solve the uploading problem, and downloading the quantum data towards classic devices could be done using classic shadow \cite{huang2020predicting}, which is exponentially efficient for sparse quantum state tomography.

In Figure~\ref{fig:qram1}, we calculate the error rate from the precision of the problem $1-F$ and the size of the data $N$, with the assumption that$T = \log N$. The region bounded by the red box indicates the current constraints used in the Supply Chain Optimization community. We use the infidelity formula proven from \cite{hann2021resilience}, $1-F \sim \frac{1}{4} \varepsilon T \log N \sim \frac{1}{4} \varepsilon \log ^2 N$ for QRAM architectures. The color density is the error rate $\varepsilon$.

In Figure~\ref{fig:qram2}, we assume $g_d=1 \text{ kHz} \times 2\pi$, $\nu = 10 \text{ MHz} \times 2\pi$, and $c_d =4.5$ which is the average of the CZ and SWAP gates inside the QRAM circuit as an estimate, in the setup of \cite{hann2021practicality} with the formula $\varepsilon \approx (\kappa +\gamma )\frac{{{c}_{d}}\pi }{2{{g}_{d}}}+{{\left( \frac{{{g}_{d}}}{\nu } \right)}^{2}}$. Here, $\kappa$ and $\gamma$ are the phonon and transmon decoherence rates, $\nu$ is the free spectral range, and $g_d$ is the direct coupling. We give an emphasis on the size of the modern Supply Chain Management problem in the red line. The color density is the error rate $\varepsilon$.

\section{Simulation Experiments}
In this section, we give precise simulation details about how to solve our Markov Decision Process using quantum computing. 

Precisely speaking, we implement the policy iteration step (Step~\ref{state:policy_evaluation}) of Algorithm~\ref{alg:qpi} in quantum computers. Step~\ref{state:policy_evaluation} is essentially solving a large linear system in \eqref{eq:q_linear_system}, and we use the variational quantum algorithms introduced in Section~\ref{subsec:variational}. Under the simplified notation, $B := (I - \gamma P^\pi)$ and $q:= Q^{\pi}$, we need to solve the following linear system
\[
Bq = r,
\]
where the dimension of $B$ is $|\mathcal{S}||\mathcal{A}|$.

\subsection{LCU Coefficients}
\begin{figure}[htbp]
\centerline{\includegraphics[width =0.35 \textwidth]{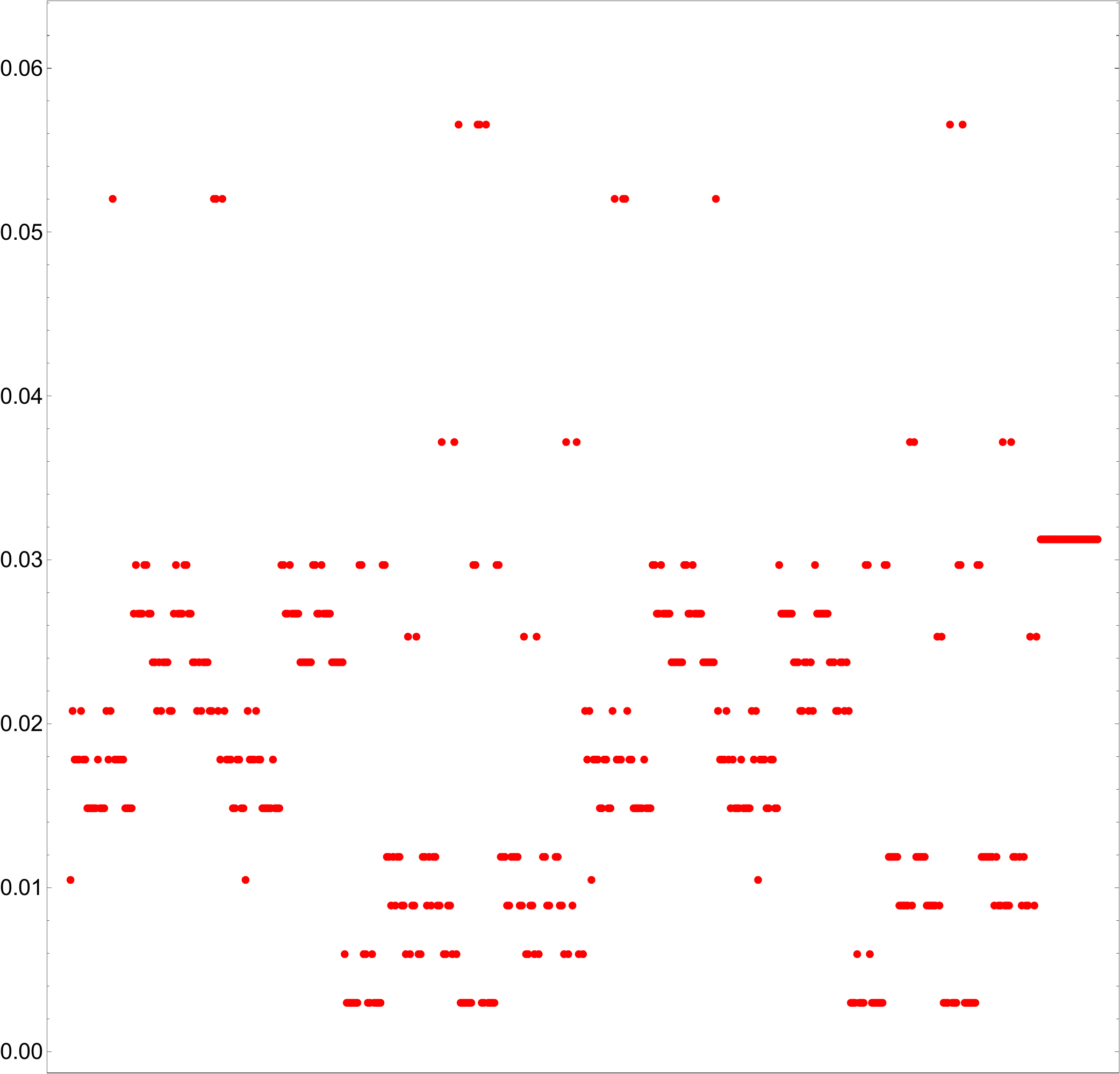}}
\caption{The distribution of the LCU coefficients of a single matrix instance used in the Markov Decision Process we study. Here we use Pauli group elements as the basis in the LCU oracle. The vertical axis is the value of the LCU coefficients, and the horizontal axis is different terms in LCU.}
\label{fig:lcu}
\end{figure}

We use the oracle model, so-called the LCU (Linear Combination of Unitaries) decomposition to upload the data of the matrix $B$ we need to invert in each iteration of the Markov Decision Process. The LCU decomposition is defined as
\begin{align}
    B = \sum_{i=1}^L a_i P_i~,
\end{align}
where $a_i$'s are real coefficients (since $B$ is Hermitian) and $P_i$s are unitary operations. One could choose $P_i$ as elements of the Pauli group ($4^N$ elements in total with $N$ qubits if we do not count for redundancies from signs), one could compute the coefficients $a_i$ according to 
\begin{align}
a_i =\frac{1}{2^N} \text{Tr}(B P_i).
\end{align}
Figure \ref{fig:lcu} shows the distribution of the LCU coefficients of a single matrix instance used in the Markov Decision Process. 

\subsection{HHL Benchmarks}
We use the \texttt{IBM Qiskit} system and the \texttt{qBraid} system to decompose the HHL unitaries towards fundamental gates. In Table \ref{table:1}, we give precise gate counting by truncating and scaling the number of LCU terms before implementing them into HHL. With our truncation and approximation scheme, we verify that the gate counting scaling is roughly polynomial and efficient, but the numbers of fundamental gates are generally high. Those results indicate that HHL is more suitable to be implemented with the help of fault-tolerant quantum computing.

\begin{table}[htbp]
\caption{Gate counting used in the HHL algorithm from real Supple Chain Optimization problems up to 6 qubits.}
\begin{center}
\begin{tabular}{ |c|c|c|c|c| } 
 \hline
  & $L=1^2$ & $L=2^2$ & $L=3^2$ & $L=4^2$ \\
 \hline
 $N=1$ & 75 & \XSolid & \XSolid & \XSolid \\
 \hline
 $N=2$ & 273 & 274 & 575 & 2215 \\ 
 \hline
 $N=3$ & 613 & 1500 & 3110 & 5086 \\ 
 \hline
 $N=4$ & 1161 & 3090 & 6466 & 10071 \\
 \hline
 $N=5$ & 2672 & 8388 & 14301 & 19329\\ 
 \hline
 $N=6$ & 2680 & 5898 & 13290 & 28143\\ 
 \hline
\end{tabular}
\end{center}
\label{table:1}
\end{table}

In Table \ref{table:1}, we compute the number of fundamental gates used in \texttt{IBM Qiskit} for the HHL oracle for the matrix inversion task in the Markov Decision Process. We use \XSolid \quad to denote the disallowed situation where the dimension of the Hilbert space cannot hold so much independent $L$. Moreover, in order to scale, we choose the leading submatrix with the corresponding size by given  number of qubits $N$ and the number of the LCU terms $L$.

\subsection{Variational Circuits}
\begin{figure}[htbp]
\centerline{\includegraphics[width =0.45 \textwidth]{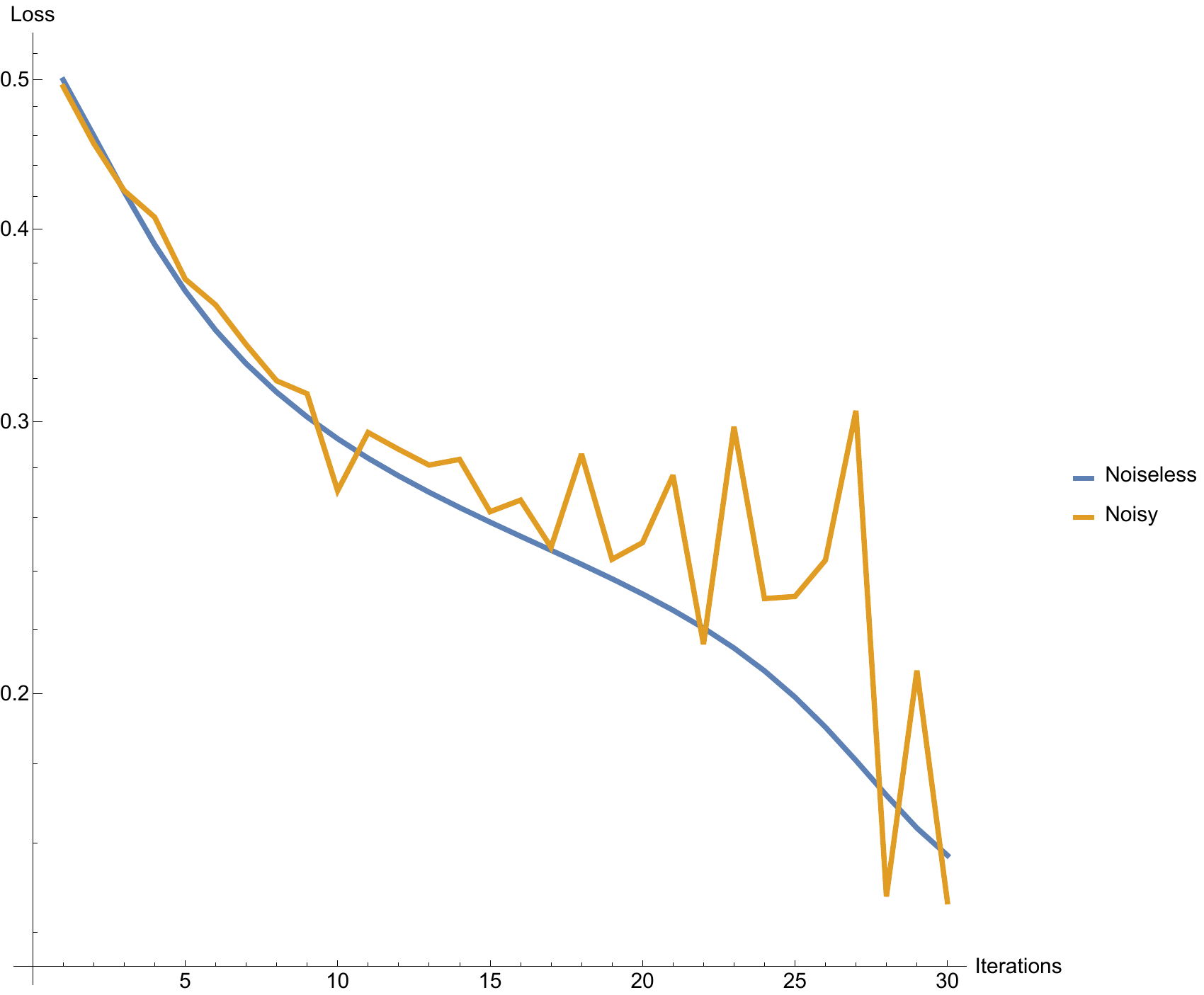}}
\caption{Solving the matrix inversion problem by variational quantum linear solver.}
\label{fig:variational}
\end{figure}

We use the \texttt{IBM Qiskit} system and the \texttt{qBraid} system to simulate the variational quantum algorithms for linear system solving, with the Markov Decision Process. With the \texttt{qBraid} environment, we compare the simulations between noiseless and noisy environments provided by \texttt{IBM Qiskit} with the real hardware noise models, with first five truncated LCU decomposition coefficients and 6 qubits, in Figure \ref{fig:variational}. We find that the variational simulation could converge in a decent amount of time even in the noisy environment. 

In Figure~\ref{fig:variational}, we use the variational ansatz provided in \cite{bravo2019variational} with 2 layers and 6 qubits. The noise calculation is from the real quantum device model in \texttt{IBM Qiskit}.

\section{Conclusion}
Despite the promising future of quantum computing, more efforts are needed to make it practical in solving real-world problems. This work explores the usage of quantum computing in the field of supply chain management by focusing on a canonical inventory control problem. We discuss in-depth  a classic inventory control problem and propose to solve it with a quantized policy iteration algorithm. Our experiments on IBM Qiskit and the qBraid system demonstrate the practicality of variational algorithms for solving small-sized inventory control problems. We believe this is a well open area that will be interesting for both academia and industry to explore further.


\section*{Acknowledgment}

We thank three anonymous reviewers for constructive and helpful feedbacks.

\bibliographystyle{ieeetr}
\bibliography{quantum.bib}

\end{document}